\documentclass[namedreferences]{solarphysics}
\usepackage[optionalrh]{spr-sola-addons} 
\usepackage{graphicx}
\usepackage{color}
\usepackage{url}             

\ifx \doiurl    \undefined \def \doiurl#1{\href{http://dx.doi.org/#1}{\textsf{DOI}}}\fi
\ifx \adsurl    \undefined \def \adsurl#1{\href{http://adsabs.harvard.edu/abs/#1}{\textsf{ADS}}}\fi
\ifx \arxivurl  \undefined \def \arxivurl#1{\href{http://arxiv.org/abs/#1}{\textsf{arXiv}}}\fi
\ifx \urlurl    \undefined \def \urlurl#1{\href{http://#1}{\textsf{#1}}}\fi

\newcommand{\aap}{    {\it Astron. Astrophys.}}

\newcommand{\apj}{    {\it Astrophys. J.}}
\newcommand{\apjl}{   {\it Astrophys. J. Lett.}}

\newcommand{\jastp}{  {\it J. Atmos. Solar-Terr. Phys.}}

\newcommand{\nat}{    {\it Nature}}

\newcommand{\solphys}{{\it Solar Phys.}}

\newcommand{\ssr}{    {\it Space Sci. Rev.}}

\begin{document}
\begin{article}
\begin{opening}

\title{Long- and Mid-Term Variations of the Soft X-ray Flare Type in Solar Cycles}

\author{I.M.~\surname{Chertok}$^{1}$\sep
                A.V.~\surname{Belov}$^{1}$}

  \runningauthor{I.M. Chertok, A.V. Belov}
 \runningtitle{Variations of Solar Flare Type}

\institute{${}^{1}$Pushkov Institute of Terrestrial Magnetism,
            Ionosphere and Radio Wave Propagation (IZMIRAN), Troitsk, Moscow, 108840 Russia,
            email: \url{ichertok@izmiran.ru}; \url{abelov@izmiran.ru}\\}

\date{Received ; accepted }

\begin{abstract}

Using data from the \textit{Geostationary Operational
Environmental Satellites} (GOES) spacecraft in the 1\,--\,8~\AA\
wavelength range for Solar Cycles 23, 24, and part of Cycles 21
and 22, we compare mean temporal parameters (rising, decay times,
duration) and the proportion of impulsive short-duration events
(SDE) and gradual long-duration events (LDE) among C- and
$\geq$M1.0-class flares. It is found that the fraction of the SDE
$\geq$M1.0-class flares (including spikes) in Cycle 24 exceeds
that in Cycle 23 in all three temporal parameters at the maximum
phase and in the decay time during the ascending cycle phase.
However, Cycles 23 and 24 barely differ in the fraction of the SDE
C-class flares. The temporal parameters of SDEs, their fraction,
and consequently the relationship between the SDE and LDE flares
do not remain constant, but they reveal regular changes within
individual cycles and during the transition from one cycle to
another. In all phases of all four cycles, these changes have the
character of pronounced, large-amplitude \textquotedblleft
\,quasi-biennial\textquotedblright\,oscillations (QBOs). In
different cycles and at the separate phases of individual cycles,
such QBOs are superimposed on various systematic trends displayed
by the analyzed temporal flare parameters. In Cycle 24, the
fraction of the SDE $\geq$M1.0-class flares from the N- and
S-hemispheres displays the most pronounced synchronous QBOs. The
QBO amplitude and general variability of the intense
$\geq$M1.0-class flares almost always markedly exceeds those of
the moderate C-class flares. The ordered quantitative and
qualitative variations of the flare type revealed in the course of
the solar cycles are discussed within the framework of the concept
that the SDE flares are associated mainly with small sunspots
(including those in developed active regions) and that small and
large sunspots behave differently during cycles and form two
distinct populations.
\end{abstract}


\end{opening}

\section{Introduction}
\label{S-introduction}

Solar flares are very diverse in many of their features, characteristics, and parameters (see, \textit{e.g.},
\opencite{Fletcher2011} for a review). For example, in the soft X-ray range, two extreme varieties of the flares are distinguished from the viewpoint of their temporal profiles. Figure 1a illustrates an example of these extreme varieties observed by the \textit{Geostationary Operational Environmental Satellites} (GOES:
\citealp{Bornmann1996}). In this case, on 20 August 1999, a gradual long-duration (LDE) M1.8-class flare occurred at first and then in the interval of a few hours a very short-duration, impulsive, more intense M9.8-class flare took place in the same active region. For convenience and by analogy with LDEs, we will refer to such and similar flares as short-duration events (SDEs). The characteristic time of LDEs is many hours, in that the decay time significantly exceeds the rise time. Therefore, the LDE abbreviation is often treated as
Long-Decay Events \citep{Kahler1977}. On the other hand, the extremely impulsive flares last a few minutes only and have a symmetric temporal profile with approximately equally short rise and decay times without additional descending  behavior. On the standard GOES three-day plots (such as Figure 1a) such SDEs look like spikes. Of course, there are many flares of intermediate character or so-called hybrid flares in which the LDE and impulsive components are observed in various combinations \citep{Svestka1989, Tomczak2015}. Many LDE flares start with an impulsive phase. As for extreme spike flares, they may occur not only separately but also at the initial phase of combined flares and during any stage of weaker background LDEs.

It is known \citep{Pallavicini1977, Ohki1983, Svestka1989} that the SDE flares are generated in isolated, compact (confined) sources as a result of the local reconnection of small-scale magnetic structures (loops) very low over active regions (ARs). The short-duration of impulsive flares is due to the fact that in them the compact energy release is followed by rapid conductive cooling. Such flares are usually accompanied by collimated jets, which sometimes can transform into narrow coronal mass ejections (CMEs). The main distinctive features of the LDE flares result from the presence of large CMEs and long-duration post-eruptive energy release that they cause. At this phase, the active region magnetosphere, strongly disturbed by a CME, relaxes to the initial state by means of the reconnection in a long, vertical current sheet that provides prolonged additional energy release. This is accompanied by the formation of more and more new high loops, composing a coronal arcade with diverging two flare ribbons at their footpoints. Because of these properties, LDEs are often referred to as eruptive or two-ribbon flares \citep{Janvier2015, Zuccarello2017}. Thus, unlike SDEs, which have mainly a local character, LDEs span practically the entire magnetosphere over an AR. This can be seen in the example of the flares on 20 August 1999 mentioned above. As the 195~\AA\
\textit{Solar and Heliospheric Observatory/Extreme ultraviolet Imaging Telescope} (SOHO/EIT) images (Figure 1b) show, the LDE flare was accompanied by a large-scale arcade and the spike flare initiated from a small compact source located outside the LDE flare ribbon.

Quite often the intense SDE or LDE flares occur in series, when within a few days one AR produces mainly uniform short-duration spike-like flares and another AR generates predominantly homologous, gradual, long-duration flares. For example, according to the GOES soft X-ray data, a series of four very similar SDEs of the M- and X-class occurred on 6\,--\,8 September 2011 in AR11283 and seven similar LDEs of the same classes originated in AR11429 during the period of 2\,--\,13 March 2012 (see Figures 1s and 2s at the site \url{www.izmiran.ru/~ichertok/FlareVariations/}). It is obvious that such a sequence of flare activity is caused by peculiarities of the magnetic structure of the corresponding ARs, which remain able for several days to produce a local low-altitude reconnection in some cases and an extensive energy release with a developed post-eruptive phase in others.

 \begin{figure} 
  \centerline{\includegraphics[width=0.95\textwidth]
   {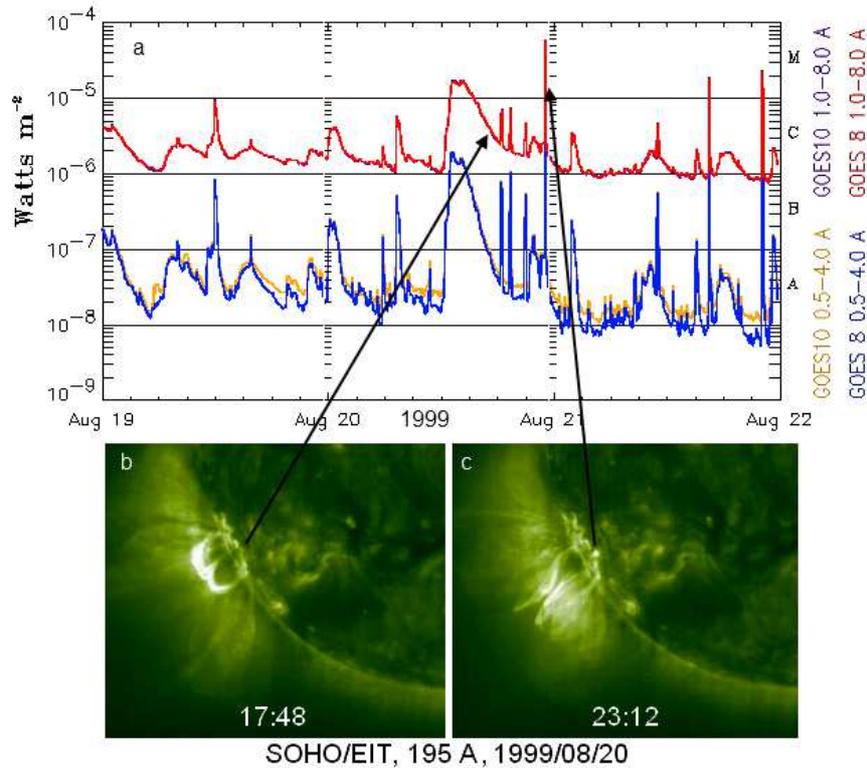}
     }
  \caption{(a) The GOES soft X-ray three-day plot illustrating the gradual LDE and impulsive SDE (spike) M-class flares of 20 August 1999. (b, c) The SOHO/EIT 195~\AA\ images displaying a post-eruptive arcade at the decay phase of the LDE flare and the compact source of the spike flare.}
  \end{figure}

Parameters of sunspots, characteristics of ARs, and solar activity as a whole strongly change in the course of the 11-year cycles (see a review of \opencite{Hathaway2015}). This stimulated us to study the cycling variations of the relationship between the SDE and LDE flares, \textit{i.e.} changes of the flare character during the recent solar cycles. An additional cause for such an analysis arose from our impression formed in a result of many-years work with the GOES soft X-ray plots, that in the current weaker Cycle 24 the fraction of the impulsive flares (including spikes) is noticeably higher in comparison with Cycle 23. In the present article, we address two issues: i) comparison of the cycles by the occurrence rate and trends of the SDE and LDE flares, which means long-term variations; ii) an analysis of the mid-term variations of the flare character during specific cycles. The structure of the article is as follows: The GOES data and analyzed parameters are described in Section 2. Subsection 3.1 is devoted to a study of histograms showing the fraction of intense and moderate SDE flares at the distinct phases of Cycles 23 and 24. In Subsection 3.2, we present the results of a more detailed analysis revealing the presence of stable, high-amplitude, antiphased quasi-biennial variations of the fraction of the SDE and LDE flares in Cycles 23, 24, and parts of in Cycles 21 and 22. In Section 4, we summarize the results and speculate that the qualitative variations of the flare character can be associated with the corresponding changes in magnetic features of sunspots and ARs in different solar cycles caused by features of the dynamo processes.


\section{Data and Parameters}
 \label{S-data}

The GOES spacecraft have provided continuous solar monitoring for more than 40 years, in particular, in the soft X-ray 0.5\,--\,4 \AA\ and 1\,--\,8 \AA\ wavelength ranges. These data are published by the U.S. National Oceanic and Atmospheric Administration (NOAA), Space Weather Prediction Center (SWPC) as the three-day plots and the flare lists indicating their coordinates, class as well as the starting, maximum, and ending times. (See \url{ftp://ftp.sec.noaa.gov/pub/warehouse/}; \url{ ftp://ftp.ngdc.noaa.gov/STP/space-weather/solar-data/solar-features/solar-flares/x-rays/goes/xrs/}, and Solar-Geophysical Data comprehensive and Weekly reports (\url{ftp://ftp.ngdc.noaa.gov/STP/SOLAR_DATA/SGD_PDFversion/})).
  In our analysis, we deal with just these temporal parameters which have been included in the IZMIRAN Database \citep{Belov2005}, in conjunction with many other solar and solar-terrestrial data. This database has allowed us to analyze the temporal characteristics of the soft X-ray flares of Cycles 21\,--\,24 in various combinations and representations.

According to the NOAA user guide (\url{ http://www.swpc.noaa.gov/sites/default/files/images/u2/Usr_guide.pdf}), the start of an X-ray flare is defined as the first minute in a sequence of four minutes of steep monotonic increase in 1\,--\,8~\AA\ flux. The time of the flare maximum is defined as the minute of the peak one-minute averaged value X-ray flux. The end time is the time when the flux level decays to a point halfway (half peak) between the maximum flux and the pre-flare background level. In comparing the SDE and LDE flares we are interested in their rising, decay times, and duration. It is clear that the rising time [$dt1$] is calculated as an interval between moments of the flare start and maximum, and the decay time [$dt2$] should be equal to the difference between the maximum and end times. Unfortunately, we are forced to accept that the flare end is determined by the flux reduction to the half-peak level. It is understood, that differences between the SDE and LDE flares would be much more obvious if the level one-
tenth, for example, were to be used instead. In such a criterion, the decay time and duration, determined as $dt = dt1+dt2$, would increase to a small degree for the SDE flares, but significantly for LDEs. Nevertheless, henceforward we will proceed from the original GOES tabular data.

We consider the sufficiently intense soft X-ray flares dividing them into two groups: i) flares of a moderate intensity with importance from C1.0 to C9.9 and ii) the most powerful flares of the $\geq$M1.0-class, including X-class flares. The reason is that the long- and medium-term variations of these intense flares can be most apparent and the peculiarities of their variations in the cycles are of most interest. Moreover, their temporal parameters, analyzed here, are not distorted in practice by the level of the background radiation, even at the maximum phase of the activity cycle.

The analysis shows (see Subsection 3.1.) that among these flares, those can be considered as impulsive or more generally as SDE ones, which have either a sufficiently rapid rise ($dt1<10$ minutes) and decay ($dt2<10$ minutes), or a short duration ($dt<15$ minutes). Below we will discuss mainly such SDE flares. In some cases they will be contrasted with long-duration flares, having particularly those with a duration $dt2>30$~minutes which we consider as LDEs.


\section{Analysis and Results}
 \label{S-analysis}


\subsection{Temporal Histograms of Flares in Cycles 23 and 24}
 \label{S-histograms}

First of all, let us compare the distributions of the flares by
their main temporal parameters in Cycles 23 and 24. We have
divided each of these cycles into three phases. For Cycles 23 and
24, respectively, the ascending phases covered during
1997\,--\,1999 and 2010\,--\,2011, the maximum phases included
2000\,--\,2002.5 and 2012\,--\,2014.5, and the declining phases
covered 2002.6\,--\,2007 and 2014.6\,--\,2016. In each of the two
cycles at these phases, the number of flares of the C1.0\,--\,C9.9
classes are measured in thousands and of the $\geq$M1.0-classes
amounted to hundreds (see the N columns in Table 1). It should be
added that the number of events at the indicated phases of Cycle
23 was more than that in Cycle 24 by factors 1.5\,--\,2.6 for the
C-class flares and factors 2.3\,--\,2.5 for the $\geq$M1.0-class
flares.

\begin{figure} 
  \centerline{\includegraphics[width=0.95\textwidth]
   {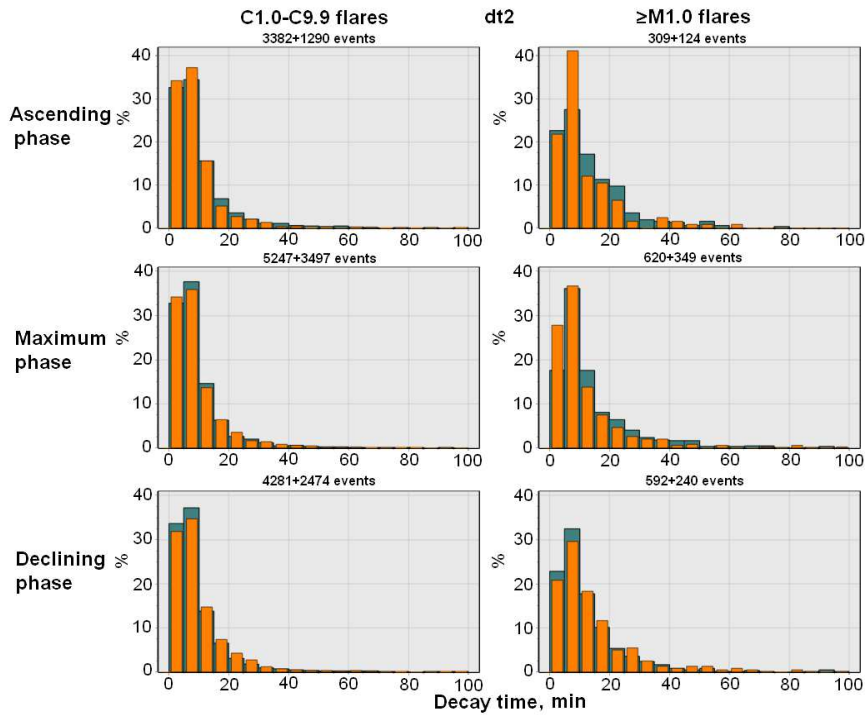}
  }
 \caption{Relative distributions (percent) of the moderate
C1.0\,--\,C9.9 class flares (\textit{left}) and the most intense
$\geq$M1.0-class flares (\textit{right}) by the decay time [$dt2$]
at the ascending, maximum and declining phases of the Cycles 23
(\textit{green}) and 24 (\textit{orange}).}
  \end{figure}

We have constructed histograms characterizing the relative
distributions of the flare numbers at the three phases of Cycles
23 and 24 according to the flare temporal parameters. Figure 2
presents one set of such histograms relating to the decay time
[$dt2$]. Here the green and orange bins belong to events of Cycles
23 and 24, and the left and right rows display the C- and the
$\geq$M1.0-class flares, respectively. It can be seen that at all
three phases of the both cycles and for two groups of flares, the
histograms are fairly similar, particularly, in the sense that
flares with the decay time $dt2$~$\approx 5-10$ minutes are
observed most often.  However, the percentage (fraction) of the
SDE flares with $dt2<10$ minutes displays some interesting
peculiarities. On the one hand, fractions of the SDE moderate
C-class flares with $dt2<10$ minutes are nearly equal at the
corresponding phases of Cycles 23 and 24: they differ by no more
than 5\,\%. On the other hand, significant differences are visible
in the powerful SDE $\geq$M1.0-class flares. The fractions of
these flares at the ascending and maximum phases of Cycle 24
exceed the same fractions in Cycle 23 by more than 10\,\%. It is
noteworthy that at the maximum phases such an excess of the
fraction of the $\geq$M1.0-class occurs due to the most SDE flares
with $dt2<5$ minutes.

The analogous histograms have been constructed for two other
temporal parameters: rising [$dt1$] and duration [$dt$]. For
brevity, they are only presented at the site mentioned above
\url{www.izmiran.ru/~ichertok/FlareVariations/}. The results of
comparison of the occurrence of the SDE flares for all three
parameters, including the  described decay time [$dt2$], are
summarized in Table 1. They confirm that Cycles 23 and 24 do not
differ in the fraction of the SDE C-class flares. By the
parameters $dt1$ and $dt$, this difference is within 2\,\%. At the
same time, for the $\geq$M1.0-class flares and these parameters, a
significant excess ($>$10\,\%) of fraction of SDEs in Cycle 24 in
comparison with Cycle 23 is found, but at the maximum phase only.

In general, the data presented in Table 1 allow us to conclude
that the relative number of the SDE\ $\geq$M1.0-class flares in
Cycle 24 exceeds that in Cycle 23 for all three temporal
parameters during the maximum phase and for the decay parameter
[$dt2$] during the ascending phase (see the bold numbers). Thus,
this corresponds to our preliminary impression resulting from the
visual inspection of the numerous GOES soft X-ray plots.


\begin{table} 
 \caption{The number [N] and percentage of the SDE C1.0\,--\,C9.9 and $\geq$M1.0 flares with $dt1$ and $dt2<$10~minutes, $dt<$15~minutes at the ascending, maximum, and declining phases of Cycles 23/24.}
 \label{T-Table1}

 \begin{tabular}{crrrrrrrr}

\hline

Cycle &  \multicolumn{4}{c}{C1.0\,--\,C9.9 flares} &
\multicolumn{4}{c}{$\geq$M1.0 flares}  \\

  phase &  \multicolumn{1}{c}{\textit{N}} & \multicolumn{1}{c}{$dt1$}   &  \multicolumn{1}{c}{$dt2$} & \multicolumn{1}{c}{$dt$} & \multicolumn{1}{c}{\textit{N}} & \multicolumn{1}{c}{$dt1$}   &  \multicolumn{1}{c}{$dt2$} & \multicolumn{1}{c}{$dt$} \\

 & \multicolumn{1}{c} { }  & \multicolumn{3}{c}{[\%]} &
 & \multicolumn{3}{c}{[\%]}  \\

\hline

Ascending  &  3382/1290 & 67/67 & 66/71  &  53/54  & 309/124  & 42/40  & 50/\textbf{62}  & 33/33 \\
Maximum  &  5247/3497 & 70/70 & 70/70  &  55/55  & 820/349  & 43/\textbf{55}  & 53/\textbf{64}  & 33/\textbf{45} \\
Declining  &  4281/2474 & 65/63 & 70/65  &  51/49  & 820/349  & 40/38  & 55/51  & 31/33 \\

\hline

 \end{tabular}

 \end{table}

\subsection{Quasi-Biennial Variations and Trends}
\label{S-biennial}

The results of the previous subsection motivated us to a more detailed analysis.  We studied further the variations of the soft X-ray flare characteristics by averaging them within the running windows of $\pm$one Carrington rotation with a step of two rotations. Near the minimum phases of the solar cycles, where the number of the C- and $\geq$M1.0-class flares is small, we kept for statistical significance only windows with $N>10$. First, consider the variations of a number of the flare parameters during Cycles 23 and 24 shown in Figure 3 against the background of the NOAA monthly sunspot number (SSN)(\url{www.ngdc.noaa.gov/stp/solar/ssndata.html}).

\begin{figure} 
  \centerline{\includegraphics[width=0.99\textwidth]
   {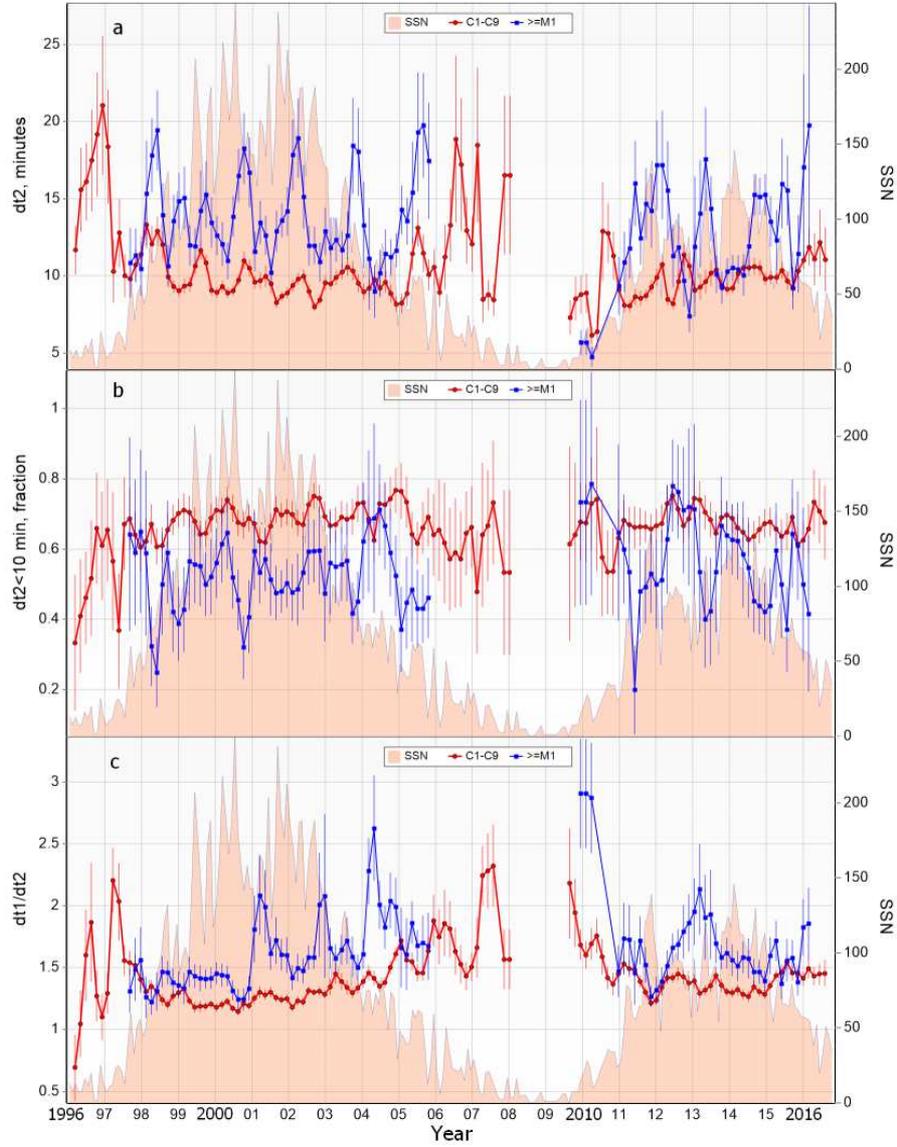}
  }
  \caption{Variations of some parameters of the C-class (\textit{red lines and circles}) and
$\geq$M1.0-class (\textit{blue lines and squares}) flares  against the background of the monthly sunspot number (SSN) during Cycles 23 and 24: (a) the average decay time [$dt2$]; (b) fraction of the SDE flares with $dt2<$10~minutes; (c) ratio of the rise and decay times [$dt1/dt2$]. Here and below the vertical line segments show $\pm 1\sigma$ standard deviations.}
  \end{figure}

Figure 3a demonstrates variations of the average decay time [$dt2$]. The first unexpected thing  is the conspicuous, very pronounced, and regular quasi-periodic variations. Their characteristic temporal ranges roughly from 0.8 to 1.9 year. Sometimes it seems to be close to the so-called Rieger-type periodicity around of 154 days, which was detected in the occurrence rate of hard-emission flares during Cycle 21 \citep{Rieger1984}. Nevertheless, following the accepted terminology (see \citealp{Bazilevskaya2014, Broomhall2015}), we will call these mid-term variations as \textquotedblleft \,quasi-biennial\textquotedblright\, oscillations (QBOs). They are observed in flares of the both C- and $\geq$M1.0-classes almost completely during Cycle 23 and 24 and have rather large amplitude, reaching
35\,--\,40~\% of the average value. Additionally it is noteworthy that the QBO amplitude of the most intense
$\geq$M1.0-class flares is almost always several times larger than that of the moderate C-class flares. Only at the beginning and end of Cycle 23 does the QBO amplitude of the C-class flares increase to the level of the $\geq$≥M1.0-class flares. Generally the QBO amplitude of the average decay time $dt2$ appears to be larger than the amplitude of its long-term changes during  both of the cycles. For most of the two cycles, the C- and $\geq$M1.0-class flares display almost synchronous variations, although sometimes this synchronism is interrupted, as  happened for example around 1998 and 2013. QBOs behave somewhat differently in the two cycles. In Cycle 23, they are more pronounced than in Cycle 24. Moreover, the average value of $dt2$, particularly of the $\geq$M1.0-class flares in Cycle 24 (12.4$\pm$0.5~minutes), is somewhat less than that in Cycle 23 (13.8$\pm$0.4~minutes). The $dt2$-parameter decreases noticeably upon transition from  Cycle 23 to Cycle 24 and reaches the smallest values at the beginning of the ascending phase of Cycle 24. During the high-activity period of Cycle 23 (1998\,--\,2005), QBOs are observed on the background of a downward trend of the $dt2$-parameter while signs of an upward trend of this parameter take place in Cycle 24.

Figure 3b shows that QBOs of the average decay time [$dt2$] described above are caused by the corresponding quasi-biennial variations of the fraction of the SDE flares with $dt2<$10 minutes. Comparison of Figures 3a and 3b shows that the $dt2$-value increases (decreases) when the fraction of SDEs decreases (rises). In particular, the fraction of the SDE flares is especially increased at the beginning of Cycle 24, and especially for the intense $\geq$M1.0-class flares. At the same time, the SDE fraction of the intense $\geq$M1.0-class flares is significantly reduced in both of the cycles during their ascending phases and during the Gnevyshev gaps between two SSN maximums of each cycle.

It is also interesting to consider variations of the ratio between the rise and decay times [$dt1/dt2$] (Figure 3c). This parameter also reveals QBOs, especially pronounced during the second half of Cycle 23. During the same part of this cycle, both the C- and ≥M1.0-class flares display an increasing trend of the $dt1/dt2$- parameter, while the opposite falling trend seems to be present during the first two years of Cycle 24 . The ratio $dt1/dt2$ is greatly increased at the beginning of Cycle 24 for the ≥M1.0-class flares, apparently because at this stage the average decay time $dt2$ was especially low for these flares  (see Figure 3a).
The same increase of $dt1/dt2$ is observed for the C-class flares, but to a lesser extent.

\begin{figure} 
  \centerline{\includegraphics[width=0.99\textwidth]
   {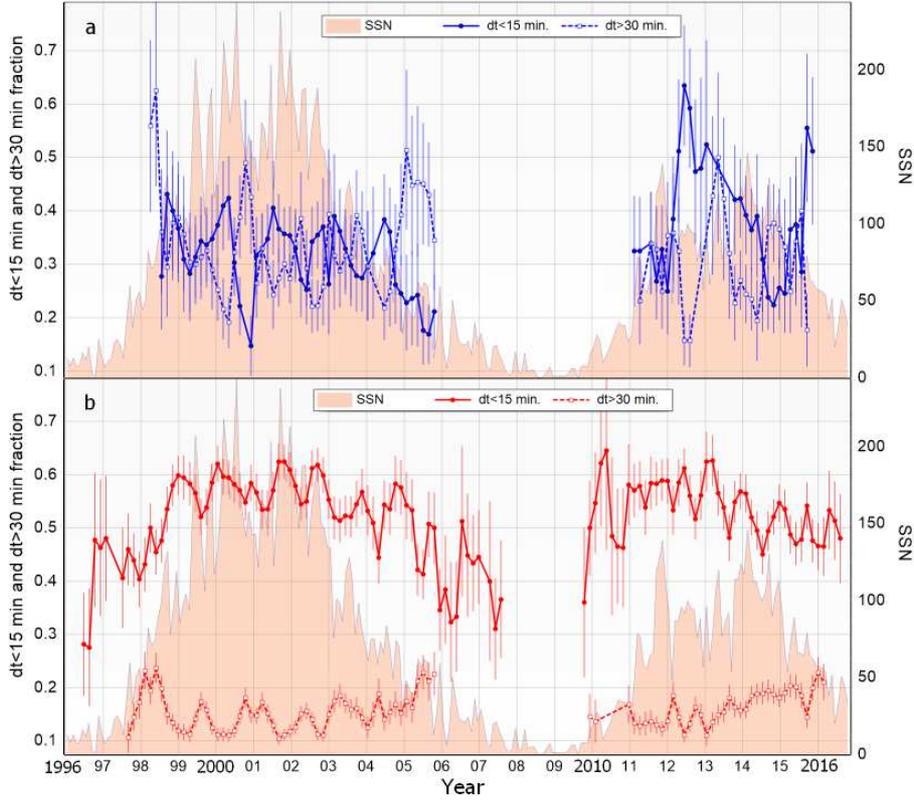}
  }
  \caption{Fraction of SDEs (\textit{solid lines}) with duration $dt<$15 minutes and LDEs (\textit{dashed lines}) with $dt>$30 minutes among the $\geq$M1.0- (a) and C-class (b) flares in Cycles 23 and 24.}
  \end{figure}

The main feature of the revealed variations outlined above is that the relationship between the SDE and LDE flares varies in a regular manner during cycles in the form of QBOs. This can be seen in Figure 4 where the fractions of SDEs with duration $dt<$15 minutes and LDEs with $dt>$30 minutes are compared for the ≥M1.0-class (a) and and C-class (b) flares.  Among the $\geq$M1.0-class flares, the fraction of such SDEs and LDEs is about the same but with some predominance of LDEs in Cycle 23 and an opposite prevalence of the SDE flares in Cycle 24. Among the C-class (b) flares, SDEs are obviously dominant. As might be expected, the relationship between SDEs and LDEs varies inversely. For the $\geq$M-class flares, opposite tendencies are observed in Cycles 23 and 24: the fraction of LDEs is increased at the onset and at the end of Cycle 23, while the fraction of SDEs is the largest after the first maximum and at the declining phase of Cycle 24. The most distinct antiphased variations of the SDE and LDE fractions took place near the first maximums and at the declining phases of the two cycles. For the C-class flares, again the SDE and LDE variability and their out-of-phase behavior is less pronounced than  for the $\geq$M-class flares.

\begin{figure} 
  \centerline{\includegraphics[width=0.99\textwidth]
   {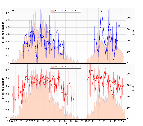}
  }
  \caption{Fraction of SDEs (\textit{solid lines}) with duration $dt<$15 minutes and LDEs (\textit{dashed lines}) with $dt>$30 minutes among the $\geq$M1.0- (a) and C-class (b) flares in Cycles 23 and 24.}
  \end{figure}

Now consider in Figure 5 the fraction of the SDE  $\geq$M1.0- (a)
and  C-class (b) flares with the duration $dt<$15 minutes occurred
separately in the north (\textit{solid line}) and south
(\textit{dashed line}) hemispheres. One can see that the clear
QBOs are also observed in this case during both Cycles 23 and 24.
However, the oscillations in Cycle 24 are much more pronounced
than in Cycle 23, especially for the intense $\geq$M1.0-class
flares in comparison with the moderate C-class flares.  The
maximum amplitude of QBOs, reaching 85\,\%, is observed in the
powerful $\geq$M1.0-class flares of the N-hemisphere near the
first maximum of the Cycle 24 in 2012. The fraction of the SDE
$\geq$M1.0-class flares decreases to the minimum values near the
Gnevyshev gaps of the both cycles and additionally at the onset of
the declining phase of Cycle 24. In general, the fractional
variations of the SDE $\geq$M1.0- and C- class flares do not
coincide in time. Another remarkable feature of Cycle 24 is a
rather high synchronicity of the fractional QBOs of the SDE
$\geq$M1.0-class flares from the N- and S-hemispheres. In Cycle
23, the similar synchronicity is observed only near the first
maximum and Gnevyshev gap. At the ascending and declining phases
of Cycle 23, the fraction of the SDE $\geq$M1.0-class flares from
the N- and S-hemispheres display mainly antiphased variations. The
intermittent synchronous and antiphased hemispherical fractional
oscillations are characteristic of the SDE C-flares flares of both
Cycles 23 and 24.

After a detailed consideration of Cycles 23 and 24, now it is
reasonable to involve data of the previous Cycles 21 and 22,
covering the period of 1977\,--\,1985 and 1987\,--\,1995  (Figure
6). However, here we confined ourselves to the rise temporal
parameter [$dt1$] because there is no certainty that the decay
time [$dt2$] and consequently duration [$dt$] were determined in
these years by the same criteria as in Cycles 23 and 24.

From Figure 6a, on which variations of the average rise time
[$dt1$] of the soft X-ray flares are presented, it follows that
QBOs are characteristic not only of Cycles 23 and 24, but also of
Cycles 21 and 22. In all four cycles, the QBO amplitude of the
strongest $\geq$M1.0-class flares considerably exceeds that of the
moderate C-class flares. Among other features, it is important to
pay attention to the growing trends of the average $dt1$-parameter
in Cycles 21\,--\,23. The smallest values of $dt1$ up to
5\,--\,6~minutes, intrinsic to SDEs, were observed at the
beginning of Cycle 21 for C-class flares and at the beginning of
Cycle 22 for $\geq$M1.0-class flares. The largest average
$dt1$~$\approx20-25$~minutes, characteristic of the
$\geq$M1.0-class LDEs, occurred at the declining phase of Cycle
23. In Cycle 24, the average $dt1$-parameter was also large, but
somewhat less than in Cycle 23.

Figure 6b allows us again to connect the variations of the average
$dt1$-value outlined above with changes of the fraction of the SDE
flares with $dt1<$10~minutes. In separate QBOs, during each cycle
and at the transition from one cycle to another, the average rise
time increases (decreases) when the fraction of the SDE flares
decreases (increases). Moreover, the fraction of the
$dt1<$10~minutes flares displays trends that are opposite to
trends of the average $dt1$-parameter visible on Figure 6a,
especially during Cycles 21\,--\,23. One can see that the
variations of the SDE fraction of the C- and $\geq$M1.0-class
flares do not always occur synchronously. The relative QBOs
amplitude of the SDE flare fraction is larger than that of the
average $dt1$, particularly for the $\geq$M1.0-class flares in
Cycles 23 and 24.

\begin{figure} 
  \centerline{\includegraphics[width=0.99\textwidth]
   {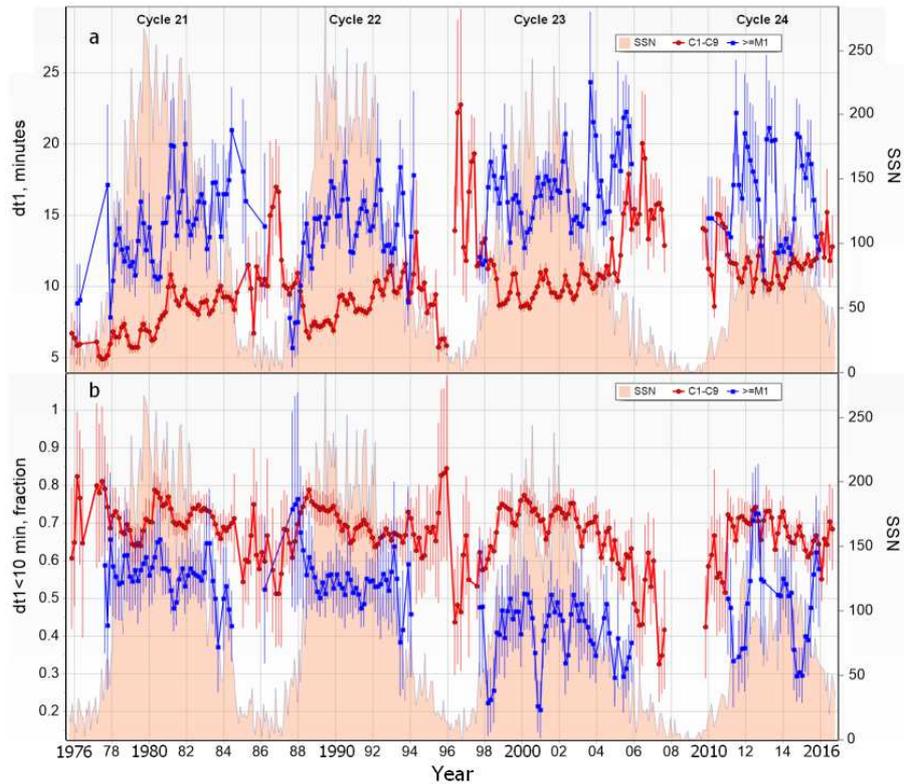}
  }
 \caption{Long-term variations of characteristics of the C-class
(\textit{red lines and circles}) and $\geq$M1.0-class
(\textit{blue lines and squares}) flares during 1976\,--\,2016
against the background of the monthly sunspot number (SSN) in
Cycles 21\,--\,24: (a) the average rise time ($dt1$); (b) fraction
of the SDE flares with $dt1<10$~minutes.}
  \end{figure}

\section{Summary and Discussion}
 \label{S-summary}

We have presented an analysis of the temporal parameters of the
soft X-ray flares registered by the GOES spacecraft during solar
Cycles 23 and 24 and additionally in Cycles 21 and 22.  We
considered the moderate C1.0\,--\,C9.9 class flares and the most
intense $\geq$M1.0-class flares and distinguished among them the
impulsive short-duration (SDE) and gradual long-duration (LDE)
flares. We focused on parameters (rising, decay times, and
duration) and fraction of SDEs and studied the relationship
between them and LDEs in the course of these cycles. Our main
results can be summarized as follows:

\begin{itemize}
\item {The fraction of the SDE $\geq$M1.0-class flares (including
spikes) in the weaker Cycle 24 exceeds that in Cycle 23 for all
three temporal parameters at the maximum phase and for the decay
parameter in the ascending cycle phase. However, Cycles 23 and 24
differ very little in fraction of the SDE C-class flares.}

\item {The temporal parameters of the SDE flares, their fraction,
and, consequently, the relationship between SDEs and LDEs do not
remain constant, but they reveal regular antiphased changes within
individual cycles and during the transition from one cycle to
another.}

\item {At all phases of the four cycles, these changes are
characterized by pronounced, large-amplitude quasi-biennial
oscillations (QBOs).  In different cycles and at the separate
phases of individual cycles, such QBOs are superimposed on various
systematic trends displayed by the analyzed temporal flare
parameters.}

\item {The QBO amplitude and general variability of the intense
$\geq$M1.0-class flares almost always markedly exceed those of the
moderate C-class flares.}

\item {The fraction of the SDE $\geq$M1.0-class flares from the N-
and S- hemispheres displays the most pronounced synchronous QBOs
in Cycle 24.}

\item {Taken together, these findings mean that ordered long- and
medium-term variations of the flare type, \textit{i.e.} alternate
transitions from the dominating SDE flares to mainly LDEs, occur
during the solar cycle.  It is important to emphasize that these
variations are not only quantitative but also qualitative in
nature.}

\end{itemize}

\medskip

As noted in the Introduction (see \opencite{Kahler1977,
Pallavicini1977, Ohki1983, Svestka1989}), LDEs are associated with
the presence of large CMEs and the long-duration post-eruptive
energy release that they cause, spanning practically the entire
magnetosphere over an AR. As opposed to LDEs, the SDE flares are
generated in isolated compact sources as a result of the local
reconnection of small-scale magnetic structures (loops) located
very low over active regions. This gives us reason to assume that
the SDE flares are associated with small sunspots, including those
in large groups.

The variations in the character of the flares found by us appear
to correspond to the concept that small and large sunspots form
two physically distinct populations and behave differently during
cycles \citep{Nagovitsyn2012, Nag-Pev2016, Nagovitsyn2016}.
According to these authors, the fraction of the smallest and
largest spots changes during the solar cycles in a systematic
antiphased way: when the number of small sunspots increases, the
number of large spots decreases and \textit{vice versa}. The
relationship between the SDE and LDE flares behaves in a similar
way. In particular, the  that the fraction of SDEs turned out to
be markedly increased at the beginning of Cycle 24 may be due to
the fact that the proportion of small spots also increased in
2010\,--\,2011 (\textit{cf.} our Figure 3 and Figure 5 in
\opencite{Nagovitsyn2012}).

As for QBOs, these medium-term variations are known to be
displayed by various solar and interplanetary proxies indicating
their relation with the global solar dynamo mechanism (see
\opencite{Bazilevskaya2014, Broomhall2015, Barlyaeva2017} and
references therein). In our case, pronounced QBOs are detected in
the temporal parameters and fraction of the SDE and LDE soft X-ray
flares. These flare QBOs differ from QBOs of other proxies by a
number of peculiarities. Their amplitude is so large (especially
for the most intense $\geq$M1.0-class flares) that their detection
does not require any spectral analysis. Also, the flare QBO
amplitude does not intensify around the cycle maxima but remains
almost unchanged and significant throughout the whole cycles. In
addition, these QBOs occur quasi-synchronously in the northern and
southern hemispheres of the Sun, particularly for the
$\geq$M1.0-class flares in Cycle 24.

The assumed connection of the regular variations of the flare
character in the course of the solar cycles, mainly in the form of
the large-amplitude QBOs, with the differing populations of small
and large sunspots has reasonable grounds. Many researchers argued
that there are two different dynamo processes that act in the deep
and near-surface layers of the convective zone and are responsible
for the 11-year cycle and more short-term variations of various
phenomena of solar activity (see \opencite{Benevolenskaya1998,
Brandenburg2005, Fletcher2010, Popova2013, Obridko2014,
Beaudoin2016}). In the context of our results, it is promising to
explore the variations of small and large sunspots using the same
approach as in the present work, i.e. averaging their number
within the running windows of $\pm$2.5 Carrington rotations with a
step of two rotations. Such advancement of the analysis of
\cite{Nagovitsyn2012} would allow a detailed comparison of the
temporal course of the SDE/LDE flares and small/large sunspots.

The detected variations of the soft X-ray flare type over the
solar cycles gives reasons to expect that similar regular
transitions between the short- and long-duration flares can occur
also in the microwave and hard X-ray ranges. This should also be a
subject of further studies.


\begin{acks}
We are grateful to the anonymous reviewer for the careful
consideration of our article. The authors thank the NOAA/SWPC GOES
and SOHO/EIT teams for the open data used in our study. SOHO is a
project of international cooperation between ESA and NASA. We are
grateful to V.V. Grechnev and V.N. Obridko for useful discussions.
This research was partially supported by the Russian Foundation of
Basic Research under grants 17-02-00308 and 17-02-00508.
\end{acks}

\medskip
\noindent
\textbf{Disclosure of Potential Conflicts of Interest} The authors have no conflicts of interest.

\end{article}

\end{document}